\documentclass[acmtog,screen,balance=true]{acmart}

\usepackage{tabularx,booktabs} %
\usepackage{multirow}
\usepackage{xcolor,colortbl,soul}
\usepackage{bm}
\usepackage{amsmath}
\usepackage{pifont}
\usepackage{mathtools}
\usepackage{wrapfig}
\usepackage{float}
\usepackage{hyperref}
\usepackage[capitalise]{cleveref}
\newcommand{\cmark}{\ding{51}}%
\newcommand{\xmark}{\ding{55}}%

\setcopyright{cc}
\setcctype{by-nc}
\acmJournal{TOG}
\acmYear{2025}
\acmVolume{44}
\acmNumber{4}
\acmArticle{}
\acmMonth{8}
\acmPrice{}
\acmDOI{10.1145/3731211}

\setcitestyle{square}

\definecolor{vermilion}{rgb}{0.89, 0.26, 0.2}
\newcommand{\new}[1]{#1}
\newcommand{\revision}[1]{#1}

\DeclareMathOperator*{\argmin}{arg\,min}

\begin{document}

\title{3DGH: 3D Head Generation with Composable Hair and Face}

\author{Chengan He}
\affiliation{%
  \institution{Yale University}
  \country{USA}
}
\email{chengan.he@yale.edu}

\author{Junxuan Li}
\affiliation{%
  \institution{Meta Codec Avatars Lab}
  \country{USA}
}
\email{junxuanli@meta.com}

\author{Tobias Kirschstein}
\affiliation{%
  \institution{Technical University of Munich}
  \country{Germany}
}
\email{tobias.kirschstein@tum.de}

\author{Artem Sevastopolsky}
\affiliation{%
  \institution{Technical University of Munich}
  \country{Germany}
}
\email{artem.sevastopolsky@gmail.com}

\author{Shunsuke Saito}
\affiliation{%
  \institution{Meta Codec Avatars Lab}
  \country{USA}
}
\email{shunsukesaito@meta.com}

\author{Qingyang Tan}
\affiliation{%
  \institution{Meta Codec Avatars Lab}
  \country{USA}
}
\email{qytan@meta.com}

\author{Javier Romero}
\affiliation{%
  \institution{Meta Codec Avatars Lab}
  \country{USA}
}
\email{javierromero1@meta.com}

\author{Chen Cao}
\affiliation{%
  \institution{Meta Codec Avatars Lab}
  \country{USA}
}
\email{chencao@meta.com}

\author{Holly Rushmeier}
\affiliation{%
  \institution{Yale University}
  \country{USA}
}
\email{holly.rushmeier@yale.edu}

\author{Giljoo Nam}
\affiliation{%
  \institution{Meta Codec Avatars Lab}
  \country{USA}
}
\email{giljoonam@meta.com}

\renewcommand{\shortauthors}{C. He et al.}

\begin{abstract}
We present 3DGH, an unconditional generative model for 3D human heads with composable hair and face components.
Unlike previous work that entangles the modeling of hair and face, we propose to separate them using a novel data representation with template-based 3D Gaussian Splatting, in which deformable hair geometry is introduced to capture the geometric variations across different hairstyles. Based on this data representation, we design a 3D GAN-based architecture with dual generators and employ a cross-attention mechanism to model the inherent correlation between hair and face. The model is trained on synthetic renderings using carefully designed objectives to stabilize training and facilitate hair-face separation. We conduct extensive experiments to validate the design choice of 3DGH, and evaluate it both qualitatively and quantitatively by comparing with several state-of-the-art 3D GAN methods, demonstrating its effectiveness in unconditional full-head image synthesis and composable 3D hairstyle editing. More details will be available on our project page: \url{https://c-he.github.io/projects/3dgh/}.

\end{abstract}



\begin{teaserfigure}
  \centering
  \includegraphics[width=\linewidth]{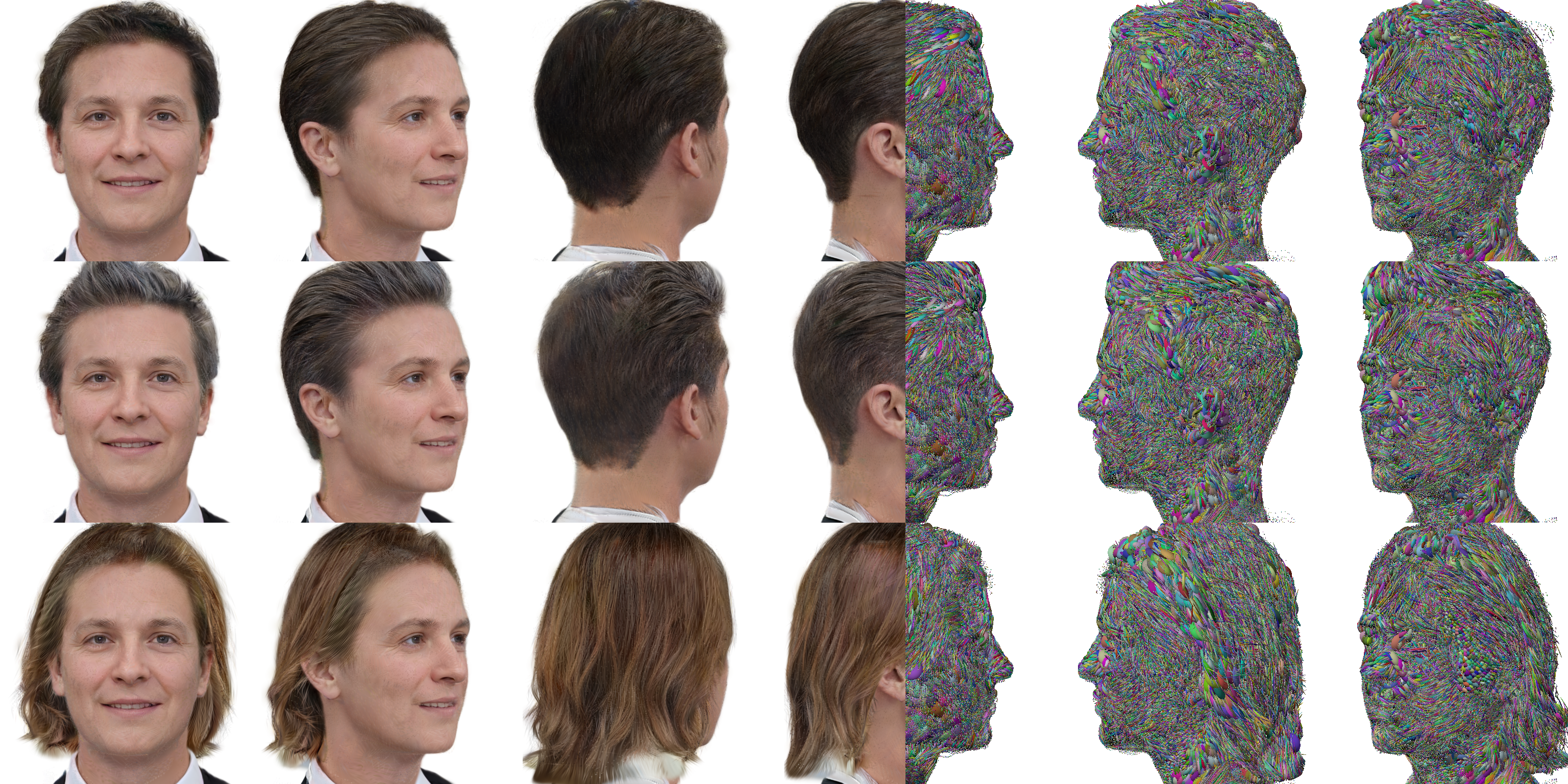}
  \caption{3DGH: Our method generates 3D head representations that can be rendered at $360^\circ$ with photorealistic rendering quality and high-fidelity geometry based on 3D Gaussian Splatting. The generated heads include composable hair and face components, enabling 3D hairstyle editing with multi-view consistency.}
  \label{fig:teaser}
\end{teaserfigure}

\maketitle
\section{Introduction}
\label{sec:intro}

The generation of high-quality 3D human heads has broad applications in digital avatars, telepresence, immersive gaming, and so on. Recently, many generative models have been proposed to facilitate 3D head generation by integrating geometry-aware representations with well-studied 2D image generative models~\cite{An_2023_CVPR,li2024spherehead,chan2022efficient,kirschstein2024gghead}.

Despite their advancements, existing methods often overlook the inherent diversity difference between hair and face, where faces of different identities still share a similarity in facial features while hairstyles are significantly more diverse. As a result, most prior 3D head generative models, which entangle the modeling of hair and face, are unsuitable for finer-grained editing tasks such as hair transfer.
Although some approaches enable editability for 2D images~\cite{nikolaev2024hairfastgan,richardson2021encoding} or geometry-aware representations like tri-plane~\cite{sun2022ide,SketchFaceNeRF2023}, they either suffer from view inconsistency in 3D applications or are inadequate for modeling 3D hair, particularly in back-of-head regions. To address these limitations, we introduce \textbf{3DGH}, a generative model for 3D full-head synthesis that supports composable hair and face components.

To train a generative model that supports compositionality, there are two key issues we need to address: (1) first, we need to ensure a clear \textbf{separation} between hair and face to disentangle these components, and meanwhile, (2) we need to respect their inherent \textbf{correlation}, as observed in real-world patterns where male faces are predominantly associated with short hairstyles, while female faces typically feature medium to long hairstyles. To tackle these challenges, we propose a novel data representation with template-based 3D Gaussian Splatting (3DGS), in which two separate mesh templates are involved to model the overall structure of hair and face with 3D Gaussians spawned on their 2D $uv$ texture maps. To capture the geometric variations among different hairstyles, we make the hair geometry itself deformable through PCA-based linear blend shapes fitted from multi-view facial capture data. This data representation then drives the design of our network architecture, which employs dual branches of StyleGAN2~\cite{karras2020analyzing} generators to independently generate hair and face Gaussians. A cross-attention mechanism~\cite{vaswani2017attention} is introduced to carefully model the inherent correlation between hair and face, ensuring coherent and realistic outputs. The model is trained using a comprehensive objective that combines adversarial loss, reconstruction terms, and regularization terms, all carefully designed to stabilize training and facilitate effective hair-face separation.

We train our model using synthetic renderings from PanoHead~\cite{An_2023_CVPR}. After training on $25m$ images, we obtain a 3D generative model capable of producing diverse 3D heads with composable hair and face components. We evaluate the model both qualitatively and quantitatively, demonstrating its effectiveness in unconditional full-head image synthesis and composable 3D hairstyle editing through comparisons with several state-of-the-art 3D GAN methods.

In summary, our contributions are as follows:
\begin{itemize}
    \item We propose 3DGH, a Gaussian-based 3D GAN for human heads that supports composable hair and face components.
    \item We introduce a novel data representation that utilizes separate template meshes for hair and face, with deformable hair geometry to capture diverse hairstyles.
    \item We design network architectures and training objectives to model hair-face separation and correlation, demonstrating their effectiveness through qualitative and quantitative comparisons with state-of-the-art 3D GAN methods.
\end{itemize}

\section{Related Work}
\label{sec:related}
In this section, we review prior work in 3D head generative models, conditional image editing methods, and 3D hair modeling approaches. 

\subsection{3D Generative Adversarial Networks}
3D Generative Adversarial Networks (GANs) are able to leverage adversarial training to develop generative models for 3D representations from 2D image collections. Early approaches primarily employed implicit 3D representations, such as NeRF~\cite{mildenhall2020nerf}, to render either raw pixels~\cite{chan2021pi, schwarz2020graf} or features subsequently decoded by a CNN-based neural renderer~\cite{niemeyer2021giraffe, xue2022giraffe}. However, the high computational cost for volume rendering posed challenges for training high-resolution GANs. To address these limitations, more recent works have adapted successful 2D GAN architectures to generate compact intermediate representations, which can then be lifted to 3D~\cite{chan2022efficient, gu2021stylenerf, or2022stylesdf}. Among these approaches, the tri-plane representation introduced in EG3D~\cite{chan2022efficient} has proven to be the most effective for generating diverse and realistic geometry-aware portrait images. Building on this foundation, subsequent works, such as PanoHead~\cite{An_2023_CVPR} and SphereHead~\cite{li2024spherehead}, extend the tri-plane representation for full-head synthesis including back-view head images.

Despite the use of compact intermediate representations, the aforementioned methods typically rely on a 2D super-resolution network to enhance efficiency during training and inference, which may introduce unwanted artifacts in the form of 3D inconsistencies as well as low-resolution geometry. 3D Gaussian Splatting (3DGS) by Kerbl et al.~\shortcite{kerbl3Dgaussians} provided an alternative direction by utilizing an explicit representation, where a set of 3D Gaussians is optimized from multi-view images using volume splatting~\cite{zwicker2001ewa}. Although the original 3DGS representation is unstructured, subsequent works such as Gaussian Shell Maps~\cite{abdal2024gaussian} and GGHead~\cite{kirschstein2024gghead} have demonstrated strategies to rig Gaussians in an organized manner relative to an underlying template mesh, thereby enabling the training of Gaussian-based 3D GANs for human body and head generations. In this work, we aim at training a similar Gaussian-based 3D GAN for human head generation, while focusing on disentangling the generation of hair and face to enable a composable model through the introduction of novel data representations, network architectures, and training strategies.

\subsection{Conditional Image Editing}

Beyond unconditional generation, GANs are extensively employed to learn mappings from a reference in a source domain to a target domain. Examples include translating semantic masks~\cite{park2019semantic, zhu2020sean} or hand-drawn sketches~\cite{chen2020deep} into photorealistic images. A common approach involves using a pre-trained StyleGAN~\cite{Karras_2019_CVPR} generator as a decoder while training customized encoders for different input modalities~\cite{richardson2021encoding}. \new{Specifically for hairstyle editing, works like~\cite{wei2022hairclip,wei2023hairclipv2,nikolaev2024hairfastgan,zhu2021barbershop} take this approach to train customized encoders to map the input conditions to the latent space of StyleGAN, thereby achieving hairstyle editing with conditions such as reference images and text}. While these methods achieve impressive results in 2D image editing, they often struggle to edit geometry-aware content due to the lack of mechanisms for preserving multi-view consistency in the synthesized outputs. To address this limitation, geometry-aware editing approaches such as IDE-3D~\cite{sun2022ide} and SketchFaceNeRF~\cite{SketchFaceNeRF2023} train 3D GANs from scratch with additional intermediate 3D representations for conditions such as semantic masks or sketches, thereby ensuring multi-view consistency during editing. \new{A concurrent work by Bilecen et al.~\shortcite{bilecen2024reference} solves a similar 3D hairstyle editing problem through tri-plane editing.} In this work, we adopt a similar approach by rendering hair-face segmentation as additional supervision. Furthermore, we carefully model the correlation between hair and face to enhance editing fidelity while respecting plausible conditional hairstyle distributions observed in the real world.

\subsection{3D Hair Modeling}

Given the complexity and variability of hair, high-quality 3D hair modeling has remained a persistent challenge for decades, as summarized in the comprehensive survey by Ward et al.~\shortcite{ward2007survey}. To capture high-quality 3D hair models, existing methods typically require high-end capture systems~\cite{paris2004capture,jakob2009capturing,xudynamichair} or even CT scanners~\cite{shen2023CT2Hair} to fully recover strand-level details, leaving them inaccessible to most users. With the availability of synthetic 3D hair datasets~\cite{hu2015single}, deep learning-based methods have emerged to regress 3D hair models from single-view~\cite{chai2016autohair,saito2018_hairvae,zhou2018hairnet,zheng2023hair,wu2022neuralhdhair} or sparse-view~\cite{zhang2017data} image input, thereby reducing the hardware requirement for 3D hair reconstruction. However, the performance of these data-driven methods is inherently constrained by the quality of their synthetic training datasets, which often lack realism and fail to represent intricate hairstyles, such as afro-textured hair. Recently, more advanced hair capture techniques have been proposed to jointly reconstruct hair geometry and appearance by incorporating neural volumetric primitives~\cite{wang2022hvh} or strand-aligned Gaussians~\cite{luo2024gaussianhair,zakharov2024gh}. While these methods produce impressive results, creating a large-scale 3D hair dataset with them remains tedious. Consequently, current hair generative models~\cite{zhou2023groomgen,sklyarova2023haar,he2025perm} continue to rely on synthetic data with augmentations and primarily focus on modeling hair geometry without appearance. In this work, we introduce a hair generative model trained on 2D image collections, capable of modeling both hair geometry and appearance with our deformable hair geometry representation and 3DGS-based rendering framework.

\section{Methodology}
\label{sec:method}

\begin{figure*}[htbp]
  \centering
  \includegraphics[width=\linewidth]{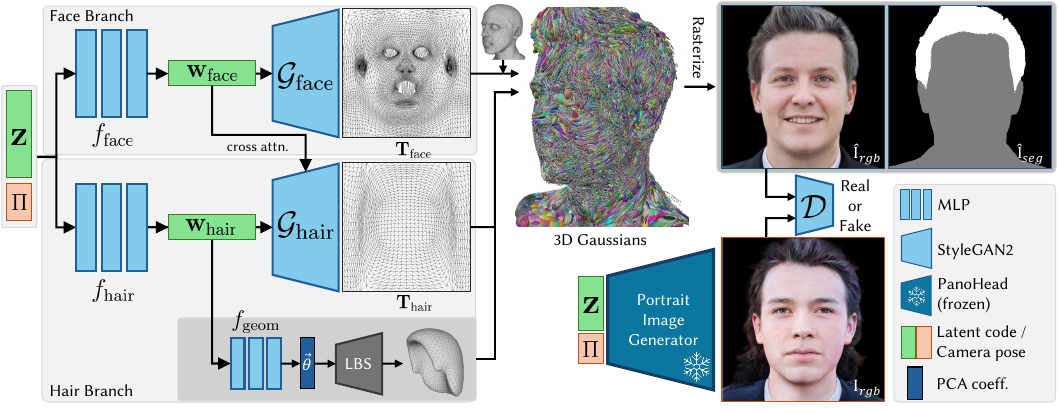}
  \caption{Overview of 3DGH, which takes a randomly sampled Gaussian noise vector $\mathbf{z}$ and camera pose $\Pi$ as input and outputs disentangled 3D head and hair representations, which are modeled as 3D Gaussians spawned on the 2D $uv$ textures of the underlying 3D meshes. The generated 3D Gaussians are rasterized and supervised by the discriminator, with additional reconstruction supervision provided by inferring PanoHead~\cite{An_2023_CVPR} using the same noise $\mathbf{z}$ and camera pose $\Pi$. In 3DGH, the hair geometric variation is modeled with a separate geometry mapping network $f_{\text{geom}}$ that produces PCA coefficients for our pre-computed linear blend shapes, and the hair-face correlation is modeled using cross-attention layers to integrate the information from $\mathbf{w}_{\text{face}}$.}
  \label{fig:overview}
\end{figure*}

An overview of our method is presented in~\cref{fig:overview}, which integrates 3D Gaussian Splatting with the well-studied 3D GAN formulation. Our approach comprises three key components: a novel data representation that incorporates 3DGS and deformable hair geometry (\cref{sec:data}), a newly designed network architecture that simultaneously models hair-face separation and correlation (\cref{sec:arch}), and training objectives specifically crafted to stabilize GAN training and enhance hair-face separation (\cref{sec:training}).

\subsection{Data Representation}
\label{sec:data}

\subsubsection{Template-Based 3D Gaussian Splatting}
Since the emergence of 3D Gaussian Splatting~\cite{kerbl3Dgaussians}, it has shown outstanding expressivity in 3D scene representation, in which the scene is represented as a collection of 3D Gaussians, with each Gaussian denoted as $\mathbf{g}_i = \{\mathbf{p}_i, \mathbf{q}_i, \mathbf{s}_i, \mathbf{c}_i, o_i\} \in \mathbb{R}^{14}$, characterized by a set of parameters. These parameters include its center position $\mathbf{p}_i \in \mathbb{R}^3$, rotation parameterized by a unit quaternion $\mathbf{q}_i \in \mathbb{R}^4$, scale factor $\mathbf{s}_i \in \mathbb{R}^3$ along each axis, color $\mathbf{c}_i \in \mathbb{R}^3$, and opacity value $o_i \in \mathbb{R}$. This collection of 3D Gaussians can then be efficiently rendered through its differentiable tile-based rasterizer given the camera pose $\Pi$.

Considering the highly unstructured nature of the original 3DGS, we follow previous works~\cite{kirschstein2024gghead,abdal2024gaussian,saito2024relightable} and associate each 3D Gaussian with a template mesh with corresponding $uv$ layout. In this way, 3D Gaussians are represented as a 2D texture map $\mathbf{T} \in \mathbb{R}^{256 \times 256 \times 14}$, where each texel stores the parameters for a single 3D Gaussian primitive. In our 3D head representation, we adopt two different meshes and texture maps for hair and face, respectively, resulting in ${\sim}131K$ Gaussians in total for the final rendering.

\subsubsection{Deformable Hair Geometry}

Even with our template-based 3DGS representation, there is an inherent difference between hair and face remaining unsolved, that is, hair contains much more geometric variation than different faces, which always share some commonalities among facial features such as eyes and mouth. For hair, there are many different hairstyles in the real world, ranging from short to long, fluffy to flat, making it hard to find a single template mesh to cover all these variations. To solve this problem, we learn a hair geometry prior that allows to produce deformable hair geometry to fit different hairstyles.

\begin{figure}[tbp]
  \centering
  \includegraphics[width=\linewidth]{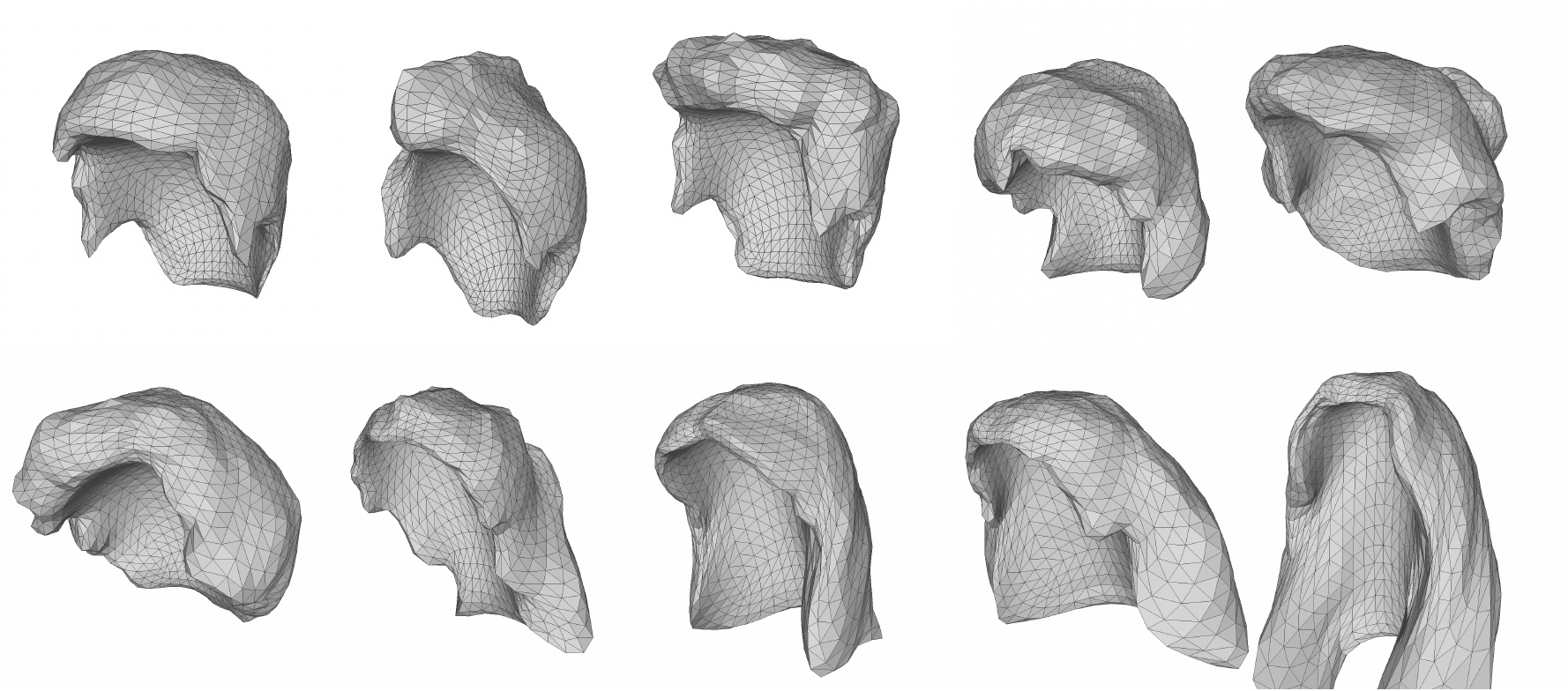}
  \caption{Examples of hair meshes fitted by our algorithm. Since all examples are deformed from the same template, they share a consistent topology, as illustrated by their wireframe visualization.}
  \label{fig:hair-geom}
\end{figure}

To learn the prior, we first fit hair geometries from multi-view facial capture data that are similar to the Multiface dataset~\cite{wuu2022multiface}, which provides calibrated camera parameters and semantic segmentations for each captured image. Given a template hair mesh, we formulate its deformation as an optimization problem, in which we differentiably render the segmentation of the deformed hair mesh using DRTK~\cite{pidhorskyi2024rasterized}, and compute the $L_1$ loss over the provided calibration. To suppress artifacts such as flipped and folded faces during deformation, we adopt an idea similar to Neural Jacobian Fields~\cite{aigerman2022neural}, where we optimize for Jacobians $\mathbf{J} \in \mathbb{R}^{F \times 3 \times 3}$ ($F$ refers to the number of faces of the template mesh) and centroid translation $\mathbf{t} \in \mathbb{R}^3$ of the template mesh, rather than vertex offsets directly. Vertex positions can be computed efficiently with a differentiable Poisson solver, and this process can thus be formally defined as:
\begin{equation}
    \mathbf{J}^\ast, \mathbf{t}^\ast \coloneqq \argmin_{\mathbf{J}, \mathbf{t}} \| \mathcal{R}\big(\text{PoissonSolve}(\mathbf{J})+\mathbf{t}; \Pi\big) - \mathbf{I}_{\text{seg}} \|_1,
\end{equation}
where $\mathcal{R}(\cdot; \Pi)$ is the differentiable rendering operator of DRTK given the camera pose $\Pi$, $\text{PoissonSolve}(\cdot)$ is the differentiable Poisson solver in Neural Jacobian Fields, and $\mathbf{I}_{\text{seg}}$ is the pre-calibrated ground truth segmentation. Our experiments show that this optimization process converges within $500$ iterations, and in \cref{fig:hair-geom} we visualize $10$ hair meshes fitted from this process.

In total, we collect $283$ different hair meshes, and then learn their prior using the conventional PCA-based methods in digital humans~\cite{blanz19993dmm, SMPL:2015}. Specifically, we solve a set of linear blend shapes by performing PCA on the normalized hair meshes, and different deformed hair meshes can thus be obtained from the linear function $\mathcal{M}(\vec\theta)$:
\begin{equation}
    \mathcal{M}(\vec\theta; \mathbf{X}) = \bar{\mathbf{M}} + \sigma\sum_{n=1}^{|\vec{\theta}|}\vec{\theta}_n\mathbf{X}_n,
\end{equation}
where $\vec{\theta} = [\theta_1, \dots, \theta_{|\vec{\theta}|}]^\top$ is a vector of blend shape coefficients, and $\mathbf{X} = [\mathbf{X}_1, \dots, \mathbf{X}_{|\vec{\theta}|}]^\top \in \mathbb{R}^{|\vec{\theta}| \times 3V}$ forms a matrix of orthogonal principal components of shape displacements, with $V$ referring to the number of vertices of the template mesh, and $\bar{\mathbf{M}} \in \mathbb{R}^{3V}$ and $\sigma \in \mathbb{R}$ denote the mean shape and standard variation computed from the original fitted hair meshes. We set the number of blend shape coefficients $|\vec{\theta}|=32$, which ensures that the deformed mesh is smooth while still covering enough variations.

\subsection{Network Architecture}
\label{sec:arch}

To obtain enough training data of frontal and back-of-head images with accurate camera poses, we adopt PanoHead~\cite{An_2023_CVPR} as our training data generator and train our generative model following the scheme of StyleGAN2~\cite{karras2020analyzing}. Formally, given a randomly sampled latent code $\mathbf{z}\in \mathbb{R}^{512}$ and camera pose $\Pi \in \mathbb{R}^{25}$, they are first passed to PanoHead to obtain a rendered RGB image $\mathbf{I}_{\text{rgb}} \in \mathbb{R}^{512 \times 512 \times 3}$, which is later segmented and parsed to obtain the foreground mask $\mathbf{I}_{\text{mask}} \in \mathbb{R}^{512 \times 512}$ and the hair-face segmentation map $\mathbf{I}_{\text{seg}} \in \mathbb{R}^{512 \times 512}$ using~\cite{lin2021real,zheng2022farl}. These images, i.e., $\mathbf{I}_{\text{rgb}}$, $\mathbf{I}_{\text{mask}}$, $\mathbf{I}_{\text{seg}}$, serve as supervision signals to train our model.

\subsubsection{Dual-Branch 3D GAN}

As illustrated in \cref{fig:overview}, we design two separate branches to handle the generation of hair and face respectively. Given the same latent code $\mathbf{z}$ and camera pose $\Pi$, the mapping network $f: \mathcal{Z} \mapsto \mathcal{W}$ is first introduced to map them to the intermediate latent space, denoted as $\mathbf{w}_{\text{hair}}$ and $\mathbf{w}_{\text{face}}$. For hair, an additional geometry mapping network $f_{\text{geom}}: \mathcal{W} \mapsto \vec{\theta}$ is designed, which maps the latent code $\mathbf{w}_{\text{hair}}$ to proper blend shape coefficients that can represent the global shape of the hairstyle. These intermediate latent codes are then fed into two separate StyleGAN generators $\mathcal{G}_{\text{hair}}$ and $\mathcal{G}_{\text{face}}$, yielding two textures $\mathbf{T}_{\text{hair}}$ and $\mathbf{T}_{\text{face}}$ that store Gaussian parameters on each texel. We spawn 3D Gaussians from these textures and associate them with the underlying template mesh, combining and rendering them together to obtain the rendered RGB image $\hat{\mathbf{I}}_{\text{rgb}}$ and mask $\hat{\mathbf{I}}_{\text{mask}}$ from the provided camera pose. We use a similar dual discrimination method as EG3D~\cite{chan2022efficient}, where we concatenate the rendered RGB and mask images and feed them into the discriminator with the camera pose $\Pi$. Aligning with the findings of Mimic3D~\cite{bib:mimic3d}, this adversarial training scheme helps increase diversity and maintain high-frequency details in our generated outputs.

\subsubsection{Hair-Face Correlation}
\begin{figure}[tbp]
  \centering
  \includegraphics[width=\linewidth]{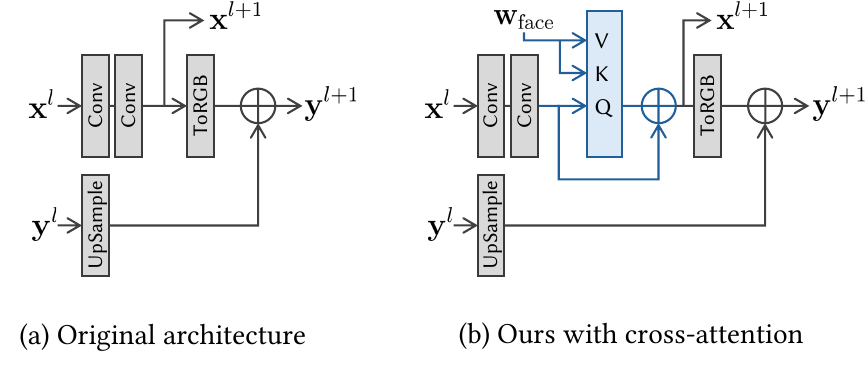}
  \caption{Diagram of our hair-face correlation module, which utilizes cross-attention layers to inject $\mathbf{w}_{\text{face}}$ into each synthesis block of StyleGAN2.}
  \label{fig:hair-face-corr}
\end{figure}

In reality, the distributions of plausible faces and hairstyles are correlated. For instance, hairstyles often correlate with gender and ethnicity. 
\revision{
To encourage our model to learn these correlations, which are commonly observed in the real world, 
}
we use cross-attention layers~\cite{vaswani2017attention} to inject $\mathbf{w}_{\text{face}}$ into each synthesis block of $\mathcal{G}_{\text{hair}}$, thereby influencing the hair generation process at different scales. Specifically, the intermediate feature map $\mathbf{y}^{l+1}$ generated at layer $l+1$ is computed as:
\begin{equation}
\label{eq:cross-attn}
    \begin{aligned}
        \mathbf{x}^{l} &= \text{Conv}(\mathbf{x}^l) \\
        \mathbf{x}^{l+1} &= \mathbf{x}^{l} + \mathcolor{vermilion}{\text{CrossAttention}(\mathbf{Q}=\mathbf{x}^{l}, \mathbf{K}=\mathbf{V}=\mathbf{w}_{\text{face}})} \\
        \mathbf{y}^{l+1} &= \text{Upsample}(\mathbf{y}^{l}) + \text{ToRGB}(\mathbf{x}^{l+1}) \\
    \end{aligned}
\end{equation}
In \cref{eq:cross-attn}, $\mathbf{x}^l$ is the input feature map for convolution from layer $l$, $\text{Conv}(\cdot)$ is the modulated convolution layers inside synthesis blocks, $\text{Upsample}(\cdot)$ is the spatial upsampling operator to upscale the previous feature map $\mathbf{y}^{l}$ to match the spatial resolution of $\mathbf{x}^{l+1}$, $\text{ToRGB}(\cdot)$ is the convolution layer that adjusts the number of channels in the convolved feature map $\mathbf{x}^{l+1}$ to match $\mathbf{y}^{l}$, and $\text{CrossAttention}(\cdot, \cdot)$ is the cross-attention layers we newly introduced compared to the original synthesis blocks in StyleGAN2. The diagram for our hair-face correlation module is provided in~\cref{fig:hair-face-corr}.

Inspired by classifier-free guidance~\cite{ho2022classifier}, we design a similar technique by randomly dropping the condition $\mathbf{w}_{\text{face}}$ (replacing it with all-zero vectors $\varnothing$) during training with a probability of $10\%$. Then in the inference stage, we blend the conditional feature map $\mathbf{x}^{l}$ and unconditional feature map $\mathbf{x}^{l}_{\varnothing}$ with the CFG factor $\omega$:
\begin{equation}
    \tilde{\mathbf{x}}^{l} = \omega\mathbf{x}^{l} + (1 - \omega)\mathbf{x}^{l}_{\varnothing},
\end{equation}
thereby allowing for further control over the hair-face correlation with the CFG factor $\omega$.

\subsection{3D GAN Training}
\label{sec:training}

As PanoHead gives us direct supervision during training, our final training objective consists of adversarial loss, reconstruction losses on rendered images, and several regularization terms to stabilize the training and improve the generation quality.

\paragraph{Adversarial Loss} Following EG3D~\cite{chan2022efficient}, we incorporate the standard non-saturating GAN loss $\mathcal{L}_{\text{adv}}$~\cite{goodfellow2014generative} with $R_1$ gradient regularization~\cite{mescheder2018training} on both the RGB and mask images, where the regularization strengths are set to $1$ for both of them.

\paragraph{RGB and Mask Loss} From Gaussian Splatting we can obtain the rendered RGB image $\hat{\mathbf{I}}_{\text{rgb}}$ and mask $\hat{\mathbf{I}}_{\text{mask}}$, on which we compute the $L_1$ loss to measure their reconstruction quality. These loss terms can be expressed as:
\begin{equation}
    \mathcal{L}_{\text{rgb}} = \|\hat{\mathbf{I}}_{\text{rgb}} - \mathbf{I}_{\text{rgb}} \|_1, \quad \mathcal{L}_{\text{mask}} = \|\hat{\mathbf{I}}_{\text{mask}} - \mathbf{I}_{\text{mask}} \|_1.
\end{equation}

\paragraph{Segmentation Loss} To encourage a clear separation between hair and face, we further assign different one-hot labels to Gaussians spawned from $\mathbf{T}_{\text{hair}}$ and $\mathbf{T}_{\text{face}}$ ($[0, 0, 1]$ for hair and $[0, 1, 0]$ for face, $[1, 0, 0]$ is left for background), which will be used to render an additional segmentation map $\hat{\mathbf{I}}_{\text{seg}} \in \mathbb{R}^{512 \times 512 \times 3}$ with other Gaussian parameters. To ensure that the generated hair mesh faithfully represents the hairstyle in the image, we additionally render the hair mesh segmentation $\hat{\mathbf{I}}_{\text{seg}}^{\text{mesh}} \in \mathbb{R}^{512 \times 512}$ with DRTK~\cite{pidhorskyi2024rasterized} by assigning different scalar values to vertices of the hair and face meshes ($2$ for hair and $1$ for face, the default value $0$ is for background). These segmentation maps are supervised with the segmentation map $\mathbf{I}_{\text{seg}}$ parsed from $\mathbf{I}_{\text{rgb}}$, where the loss terms are defined as:
\begin{equation}
    \mathcal{L}_{\text{seg}} = \text{CrossEntropy}(\hat{\mathbf{I}}_{\text{seg}}, \mathbf{I}_{\text{seg}}), \quad \mathcal{L}^{\text{mesh}}_{\text{seg}} = \|\hat{\mathbf{I}}^{\text{mesh}}_{\text{seg}} - \mathbf{I}_{\text{seg}} \|_1.
\end{equation}
Note that we use the cross-entropy loss rather than $L_1$ loss for the segmentation map $\hat{\mathbf{I}}_{\text{seg}}$ rendered from Gaussian Splatting, as its $\alpha$-blending nature will inevitably change values around the boundary, thus causing mislabeling issues for pixels on the boundary if their labels are scalar.

\paragraph{Regularization Terms} As our adversarial training is weakly supervised and 3D Gaussians are quite sensitive to gradient updates during early training stages, unconstrained training will quickly lead to divergence or mode collapse with overly large or extremely small Gaussians in the early stage. Therefore, we apply some regularization terms to stabilize the training. First, based on our hybrid 3DGS representation, the center position $\mathbf{p}_i$ of each Gaussian is defined as the sum of the 3D position $\mathbf{v}_i$ on the mesh surface and the delta position $\Delta\mathbf{p}_i$ decoded from the generated texture maps. To ensure that all Gaussians stay within a thin layer around the mesh surface, we first clamp the absolute value of the decoded delta position $\Delta\mathbf{p}_i$ with a threshold $\gamma$. We use $\gamma=40$ for face Gaussians, meaning that they can move at most $40$mm away from the surface. As the hair mesh itself is deformable, we reduce $\gamma$ to $20$ for hair Gaussians to make sure that different hairstyles are generated with different geometries, rather than similar geometries with largely deviated Gaussians.
A regularization term for delta positions is applied:
\begin{equation}
    \mathcal{L}^{\text{pos}}_{\text{reg}} = \sum_{i}\|\Delta\mathbf{p}_i\|_2,
\end{equation}
which encourages all predicted Gaussians to stay close to the mesh surface. To constrain the scale of Gaussians to stay within a reasonable range, the most effective regularization term we experimented with can be defined as:
\begin{equation}
    \mathcal{L}^{\text{scale}}_{\text{reg}} = 
    \begin{cases}
        10\times|s_i - s_{\text{min}}| & s_i < s_{\text{min}} \\
        (s_i - s_{\text{max}})^2 & s_i > s_{\text{max}} \\
    \end{cases}
\end{equation}
It applies different penalties to constrain the Gaussian scaling along all axes if they are outside of a reasonable range $[s_{\text{min}}, s_{\text{max}}]$, where $s_{\text{min}}=0.2$ and $s_{\text{max}}=5$. Finally, we also apply the $uv$ total variation loss $\mathcal{L}^{\text{uv}}_{\text{reg}}$ proposed in GGHead~\cite{kirschstein2024gghead} to prevent Gaussians in the back from shining through to the front.

Combining all the terms discussed above, the final training objective is defined as their weighted sum,
expressed as:
\begin{equation}
    \begin{aligned}
        \mathcal{L} &= \mathcal{L}_{\text{adv}} + \lambda_{\text{rgb}} \mathcal{L}_{\text{rgb}} + \lambda_{\text{mask}} \mathcal{L}_{\text{mask}} + \lambda_{\text{seg}} \mathcal{L}_{\text{seg}} + \lambda^{\text{mesh}}_{\text{seg}} \mathcal{L}^{\text{mesh}}_{\text{seg}} \\
        &+ \lambda^{\text{pos}}_{\text{reg}} \mathcal{L}^{\text{pos}}_{\text{reg}} + \lambda^{\text{scale}}_{\text{reg}} \mathcal{L}^{\text{scale}}_{\text{reg}} + \lambda^{\text{uv}}_{\text{reg}} \mathcal{L}^{\text{uv}}_{\text{reg}}
    \end{aligned}
\end{equation}
where we set the weighting factors $\lambda_{\text{rgb}}=10$, $\lambda_{\text{mask}}=10$, $\lambda_{\text{seg}}=1$, $\lambda^{\text{mesh}}_{\text{seg}}=100$, $\lambda^{\text{pos}}_{\text{reg}}=0.1$, $\lambda^{\text{scale}}_{\text{reg}}=1$, and $\lambda^{\text{uv}}_{\text{reg}}=1$ to balance the influence of different terms.

\section{Experiments}
\label{sec:experiments}

We use multi-view capture data~\cite{wuu2022multiface,saito2024relightable} to solve the linear blend shapes for our deformable hair geometry, and we use PanoHead~\cite{An_2023_CVPR} as the portrait image generator to train our generative model. For details about these datasets, please refer to Sec. A in supplemental.

\subsection{Comparisons}

We compare against several competitive baseline methods from the 3D GAN literature. Unless otherwise indicated, all baselines are their official checkpoints to maintain their original quality. We evaluate the quality of the generated multi-view images, both quantitatively and qualitatively.

\subsubsection{Qualitative Comparisons}

\begin{figure*}[ht]
  \centering
  \includegraphics[width=\linewidth]{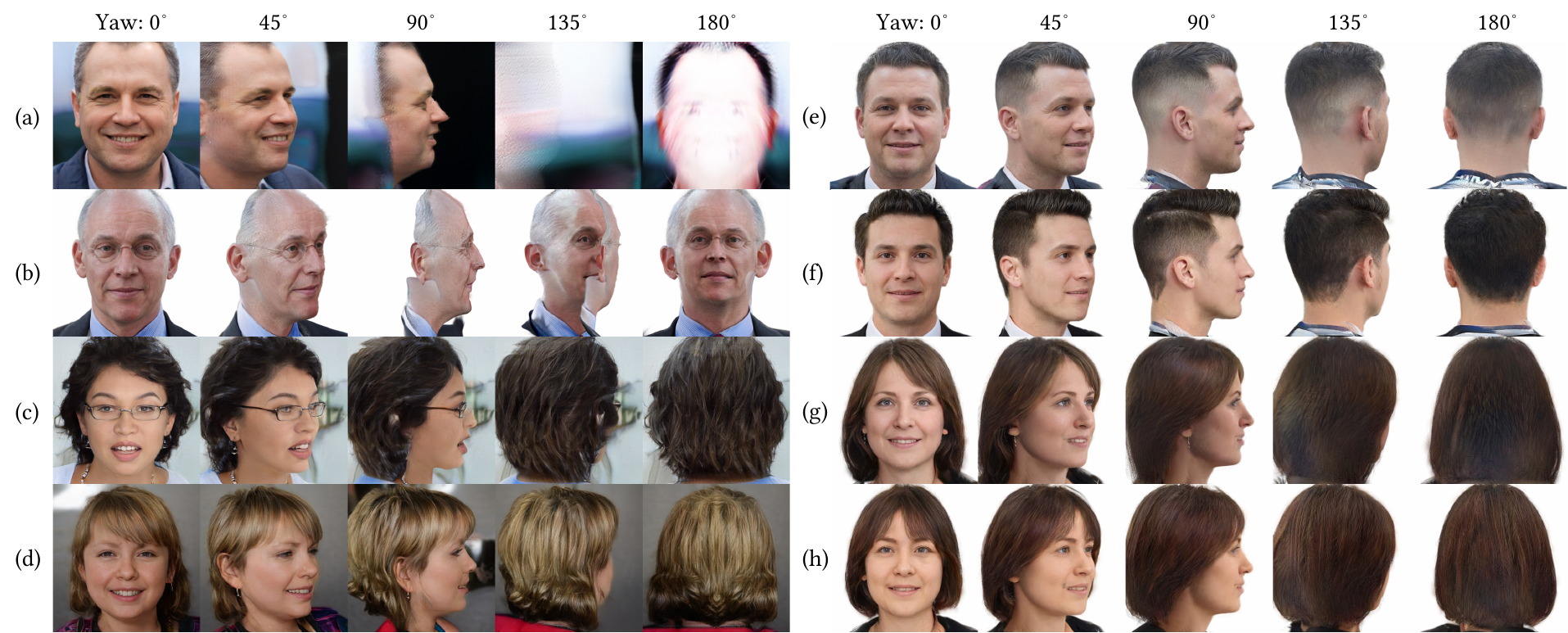}
  \caption{Qualitative comparison with various 3D GANs. (a) EG3D~\cite{chan2022efficient} and (b) GGHead~\cite{kirschstein2024gghead} are not specifically designed for full-head image synthesis, resulting in significant quality degradation when rendering large-pose or back-view images. (c) SphereHead~\cite{li2024spherehead} and (d) PanoHead~\cite{An_2023_CVPR} are two 3D GANs tailored for full-head image synthesis. Compared with these methods, our results (e-h) demonstrate comparable quality while offering the additional advantage of compositionality.}
  \label{fig:gan-baseline}
\end{figure*}

\cref{fig:gan-baseline} visually compares the image quality against baselines including EG3D~\cite{chan2022efficient}, PanoHead~\cite{An_2023_CVPR}, SphereHead~\cite{li2024spherehead}, and GGHead \cite{kirschstein2024gghead}, where EG3D is the pioneering work that synthesizes high-quality portrait images with the tri-plane representation and a 2D super-resolution network, PanoHead and SphereHead are two subsequent works that achieve full-head synthesis by improving the tri-plane representation, and GGHead utilizes a similar hybrid representation with 3DGS and a template mesh as ours. All synthesized images contain $5$ different views, with yaw angles ranging from $0^{\circ}$ to $180^{\circ}$. While all methods successfully synthesize realistic frontal-view images, the rendering quality of EG3D and GGHead deteriorates significantly when rendering from large camera poses or back-view areas, as these methods are not specifically designed for full-head image synthesis. Compared to SphereHead and PanoHead, our method achieves comparable visual quality while providing the additional advantage of compositionality.

\begin{figure*}[ht]
  \centering
  \includegraphics[width=\linewidth]{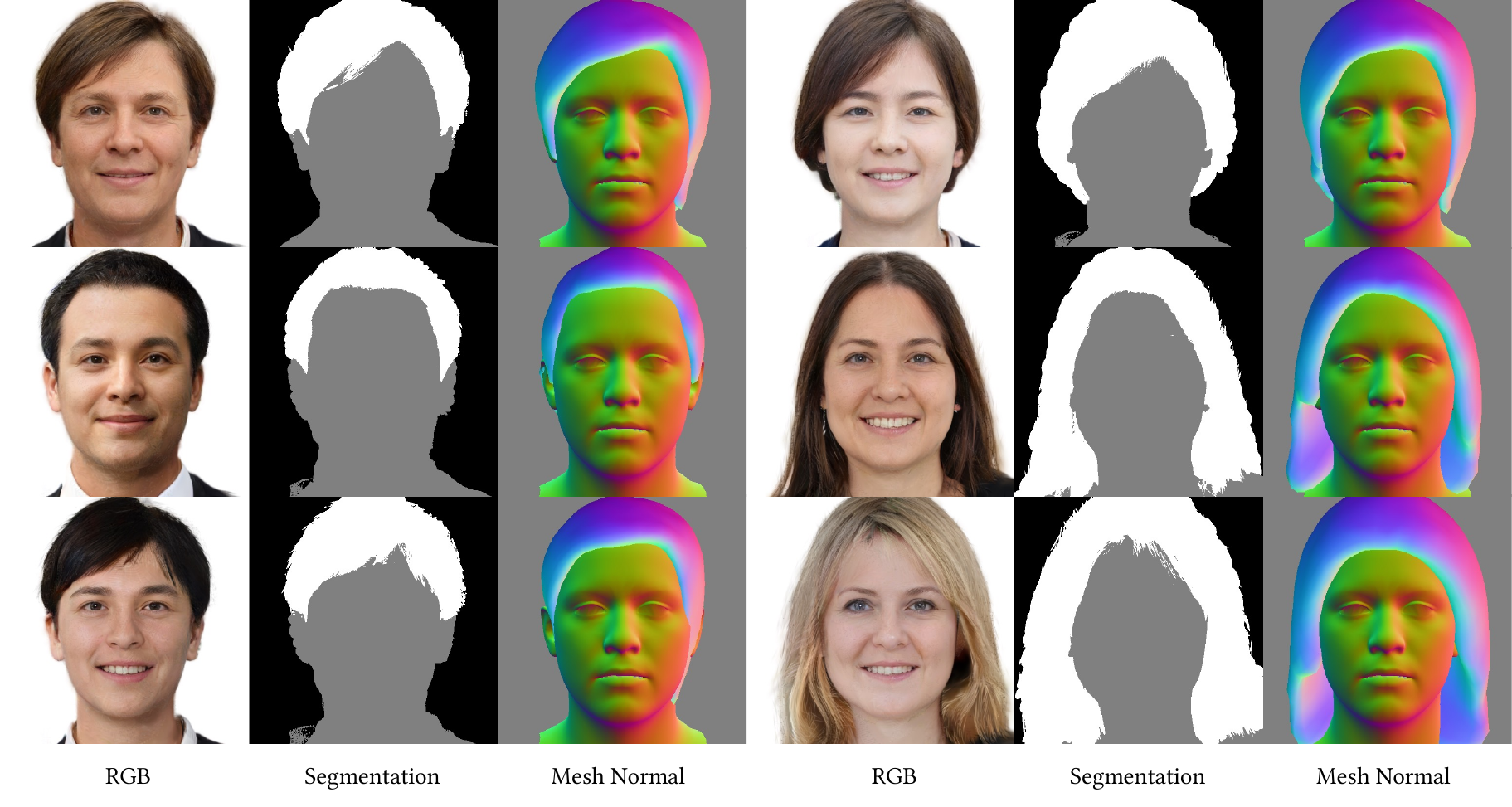}
  \caption{Generated samples with corresponding hair-face segmentation and deformed hair geometry. Our method enables smooth deformation of the hair mesh to represent various hairstyles, while assigning 3D Gaussians to capture strand-level details and appearance.}
  \label{fig:geometry}
\end{figure*}

To prove the compositionality of our method, we first present samples generated by our model alongside the rendered hair-face segmentation maps and mesh normal maps in~\cref{fig:geometry}. Leveraging our deformable hair geometry, we achieve smooth deformations of the hair mesh, effectively capturing the overall structure of various hairstyles. Additionally, the 3D Gaussians associated with the hair mesh exhibit a clear separation from the face Gaussians and are capable of representing some strand-level details, thus yielding hair-face segmentation maps with finer-grained details that are difficult to obtain from common image segmentation models. More uncurated samples can be found in Sec. D of our supplemental.

\begin{figure*}[ht]
    \centering
    \addtolength{\tabcolsep}{-3.5pt}
    \begin{tabular}{cc}
      Original Hairstyle &  \\
      \includegraphics[width=0.825\linewidth]{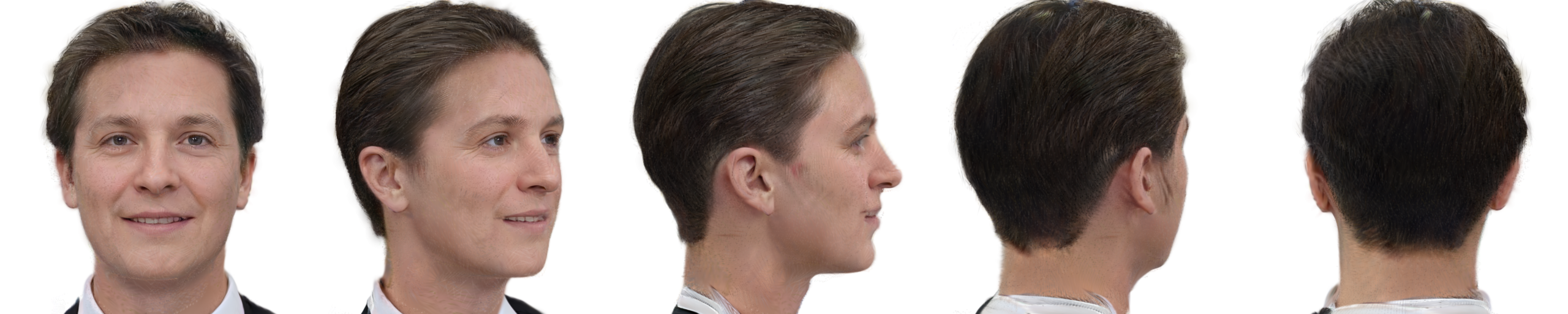} & \\
      Edited Hairstyle & Reference \\
      \includegraphics[width=0.825\linewidth]{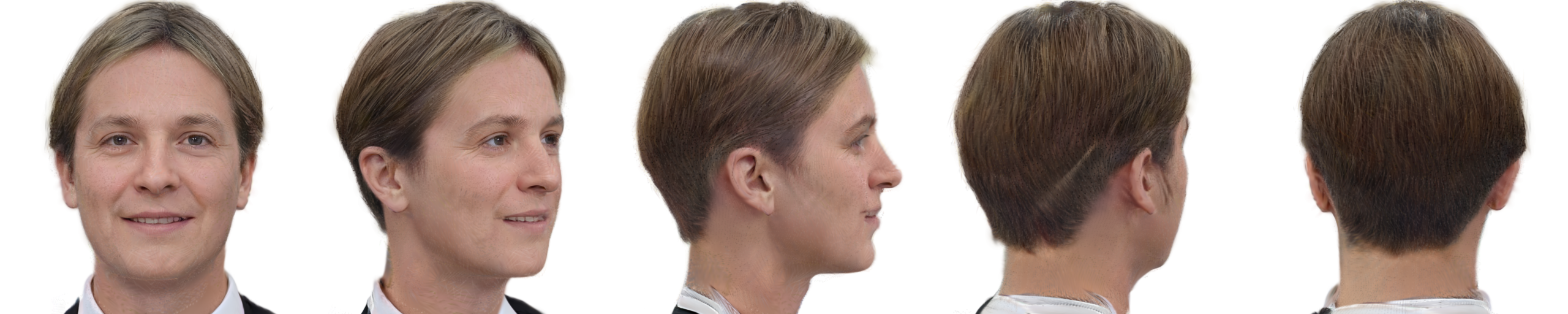} & \includegraphics[width=0.165\linewidth]{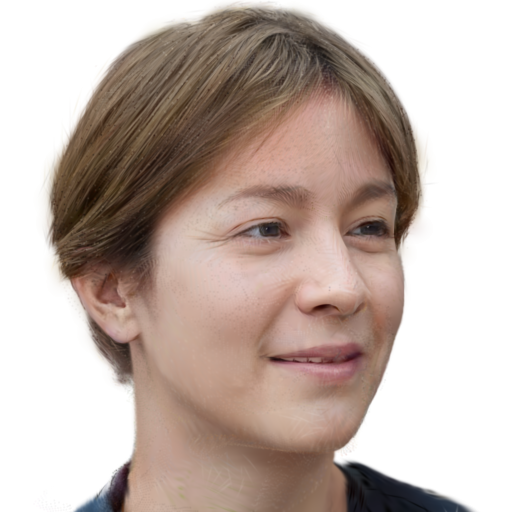} \\
      \includegraphics[width=0.825\linewidth]{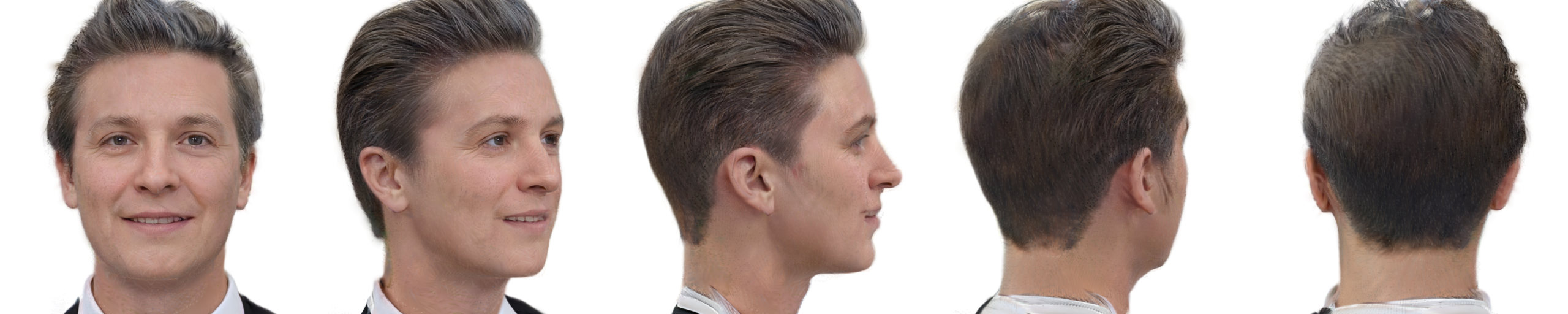} & \includegraphics[width=0.165\linewidth]{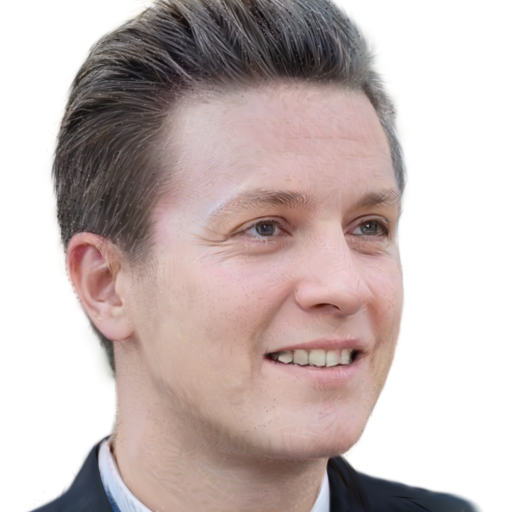} \\
      \includegraphics[width=0.825\linewidth]{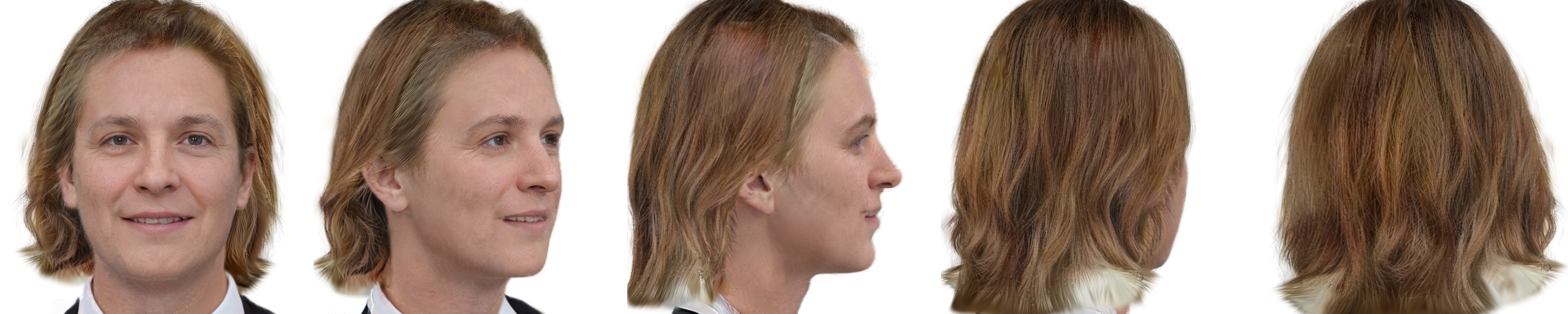} & \includegraphics[width=0.165\linewidth]{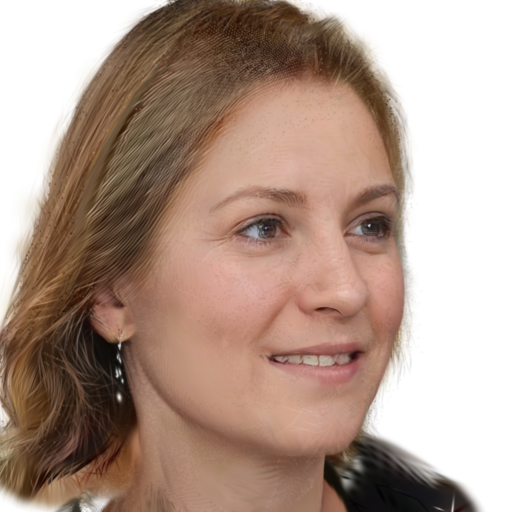} \\
    \end{tabular}
    \caption{Our method supports 3D hairstyle editing by swapping the hair latent code $\mathbf{w}_{\text{hair}}$ from the reference samples. This editing process transfers both the hair geometry and appearance, while ensuring multi-view consistency thanks to the inherently 3D nature of our representations.}
    \label{fig:edit-baseline}
\end{figure*}

We then evaluate our compositionality through 3D hairstyle editing, where the 3D hairstyle in the reference sample is transferred to another one. The qualitative results are provided in~\cref{fig:edit-baseline}, demonstrating that our approach preserves the reference hairstyle with high fidelity while producing a natural blending around the hairline. 
Both the hair geometry and appearance are transferred through a simple latent code swap. Note that in 2D hairstyle editing methods, such as HairFastGAN~\cite{nikolaev2024hairfastgan}, such hairstyle editing typically requires multiple processing steps and network modules. 
Furthermore, our editing results inherently maintain multi-view consistency, attributed to the 3D nature of our representations.

\begin{figure*}[ht]
    \centering
    \addtolength{\tabcolsep}{-2.7pt}
    \begin{tabular}{ccccc}
      \includegraphics[height=0.147\linewidth]{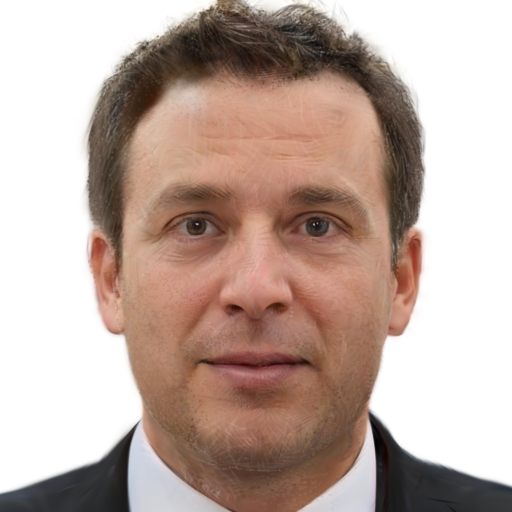} & 
      \includegraphics[height=0.147\linewidth]{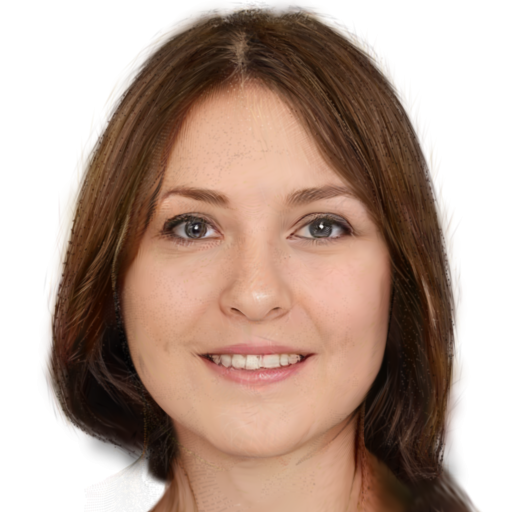} &
      \includegraphics[height=0.147\linewidth]{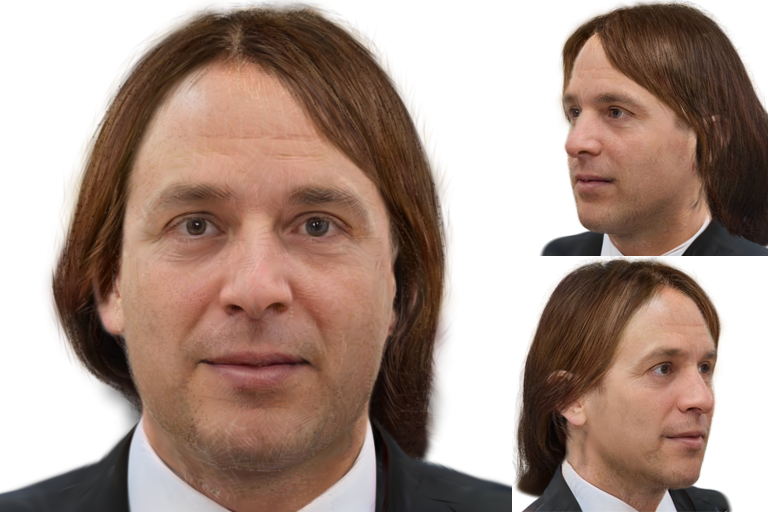} &
      \includegraphics[height=0.147\linewidth]{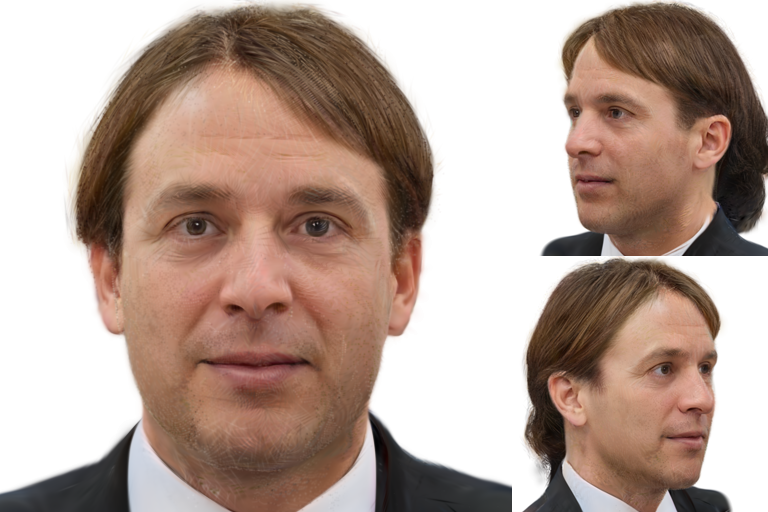} &
      \includegraphics[height=0.147\linewidth]{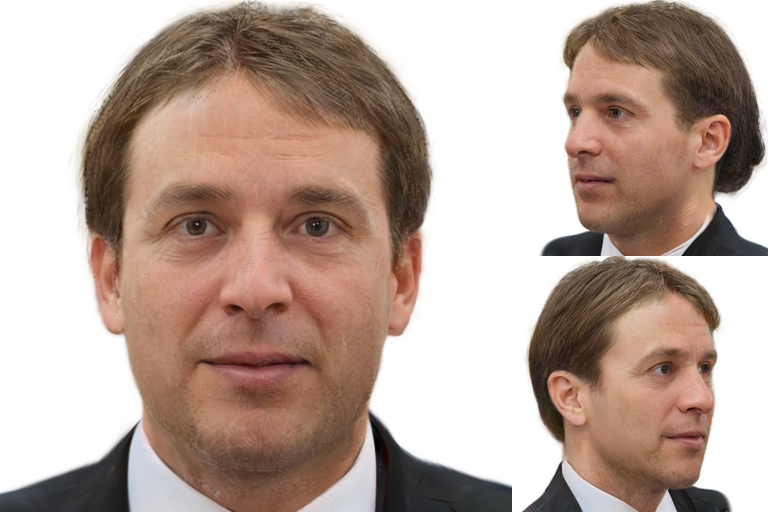} \\
      \small Reference Face & \small Reference Hair & \small $\omega=0$ & \small $\omega=0.5$ & \small $\omega=1$ \\
    \end{tabular}
    \caption{Analysis of the CFG scale factor $\omega$, where we present hair-face compositions generated using varying levels of $\omega$. When $\omega$ is small, the hair-face correlation has weak influence on the final output, resulting in hairstyles more similar to the reference. As $\omega$ increases,  the composition process becomes more biased toward the face distribution, producing hairstyles that are more contextually appropriate for the given face.}
    \label{fig:cfg-scale}
\end{figure*}

We finally investigate the impact of our hair-face correlation modeling technique by generating hair-face compositions with varying levels of the CFG scale factor $\omega$. As illustrated in~\cref{fig:cfg-scale}, when the reference face is male, it favors short-length hairstyles to align with plausible hairstyle distributions observed for this face condition in real life. Consequently, as $\omega$ increases, the transferred hairstyle gradually becomes shorter while preserving the overall style of the reference. 
This experiment provides strong evidence for the effectiveness of our hair-face correlation module, which introduces an extra dimension for editing transferred hairstyles while maintaining plausibility -- a critical point often overlooked by previous hairstyle editing methods.

\subsubsection{Quantitative Comparisons}

\begin{table}[ht]
\centering
\caption{Quantitative comparison on different FID metrics.}
\label{tab:fid}
\begin{tabular}{@{}lccc@{}}
\toprule
 & \multicolumn{3}{c}{FID $\downarrow$} \\
\cmidrule{2-4}
     & back      & front      & all     \\
\midrule
Ours & $9.86$    & $5.47$     & $6.55$  \\
\bottomrule
\end{tabular}
\end{table}

We first quantitatively evaluate our generation quality by measuring the Fr\'{e}chet Inception Distance (FID)~\cite{heusel2017gans} computed over $50k$ real and fake image samples. Since PanoHead~\cite{An_2023_CVPR} serves as our training data generator, its renderings are treated as real image samples for the FID evaluation. In addition to the overall FID (FID-all) computed on randomly sampled poses, we evaluate generation quality at a finer granularity by sampling camera poses from different regions. Specifically, FID-front evaluates facial details using images synthesized from frontal views ($|yaw| < 90^\circ$), while FID-back assesses hair details using images synthesized from back views ($|yaw| \geq 90^\circ$). As shown in~\cref{tab:fid}, all three metrics yield values less than $10$, indicating that the quality of our generated results is still comparable to that of PanoHead.

We then conduct quantitative comparisons with EG3D~\cite{chan2022efficient}, SphereHead~\cite{li2024spherehead}, and GGHead~\cite{kirschstein2024gghead} to evaluate multi-view consistency, in which we measure the identity similarity score (ID) by calculating the average Adaface~\cite{kim2022adaface} cosine similarity between paired images rendered from different camera poses.
As reported in \cref{tab:multi-view}, our method achieves the best multi-view consistency, since EG3D and GGHead struggle for large-pose images and SphereHead involves a 2D super-resolution network that may introduce artifacts.

\begin{table}[t]
\centering
\caption{Quantitative comparison between our method and other 3D GANs on multi-view consistency.}
\label{tab:multi-view}
\resizebox{\columnwidth}{!}{%
\begin{tabular}{@{}lcccc@{}}
\toprule
              & EG3D                        & GGHead                       & SphereHead              & \multirow{2}{*}{Ours} \\
              & \cite{chan2022efficient}    & \cite{kirschstein2024gghead} & \cite{li2024spherehead} & \\
\midrule
ID $\uparrow$ & $0.678$                     & $0.683$                      & $0.581$                 & $\mathbf{0.690}$ \\
\bottomrule
\end{tabular}%
}
\end{table}

\subsection{Ablation Study}

In \cref{tab:ablation}, we analyze key design decisions in our method, including the choice of supervision on segmentation maps, the use of deformable hair geometry, and variations in hair-face correlation modules. To assess the compositionality of our approach, we generate $50k$ samples by randomly swapping the intermediate latent codes $\mathbf{w}_{\text{hair}}$ and $\mathbf{w}_{\text{face}}$, thereby creating novel samples with mismatched hair and face combinations. We then compute FID for these swapped samples against the real samples, denoted as FID-swap, to evaluate the realism of the randomly combined hair and face outputs produced by our model.

In \cref{tab:ablation}, the first $3$ rows evaluate the impact of our choice of segmentation supervision. Specifically, \emph{Seg. in $\mathcal{D}$} refers to concatenate the rendered segmentation maps to the input of discriminator and let it determine whether the segmentation is realistic or not in an adversarial manner. Meanwhile, \emph{w/o Seg. loss} denotes setting $\lambda_{\text{seg}}$ to $0$. In these experiments, we observed that passing segmentation maps to the discriminator often caused mode collapse during the early stages of GAN training, resulting in meaningless outputs and high FID scores. 
\revision{We hypothesize that it is mainly due to the mismatch in value representations: the rendered segmentation maps contain continuous floating-point values, whereas the ground-truth segmentations parsed from RGB images are discrete labels. This is a fundamental difference between segmentation maps and RGB/mask images, since both ground truth RGB and mask images can contain continuous values between 0 and 1. Therefore, the discriminator can easily distinguish real and generated samples based on this quantization discrepancy, breaking the training process at an early stage.}
Surprisingly, removing the segmentation loss still produced reasonable generation results with acceptable segmentation quality. Adding the segmentation loss encouraged a cleaner separation between hair and face Gaussians, though it slightly increased the FID score due to the additional constraints imposed on Gaussians. We provide qualitative comparisons to discuss these observations in Fig. 1 of our supplemental. The $3$rd row further illustrates the importance of our deformable hair geometry. In this experiment, we replaced the deformable hair geometry with the average hair mesh fitted from studio capture data and trained our model using this fixed geometry. Quantitative results demonstrate that incorporating a deformable hair geometry improves overall generation quality. Visually, we observed that when the hair geometry is fixed, hair Gaussians need larger deviations to represent varying hairstyles, resulting in floating Gaussians appearing in random positions. Fig. 2 in supplemental includes a qualitative comparison of this artifact. The last $3$ rows examine different hair-face correlation modules. In the $4$th row, we remove this module entirely, while the $5$th row replaces it with a mechanism that concatenates $\mathbf{w}_{\text{hair}}$ and $\mathbf{w}_{\text{face}}$ with a lightweight MLP to fuse them and model their correlation. The results indicate that our cross-attention mechanism achieves the lowest FID, signifying better generation quality. Although the concatenation mechanism achieves a lower FID-swap, qualitative analysis reveals that it introduces a strong dependency on $\mathbf{w}_{\text{face}}$, which reduces the diversity when $\mathbf{w}_{\text{face}}$ is fixed and $\mathbf{w}_{\text{hair}}$ is swapped. Fig. 3 in supplemental shows this artifact. Overall, these experiments demonstrate that our final architecture design (as shown in the last row) achieves the best balance of generation quality and diversity, enabling 3D hairstyle editing in the generated results with a certain guarantee of realism.

\begin{table}[t]
\caption{Ablation studies on segmentation supervision, deformable hair geometry, and hair-face correlation module. The last row refers to our final architecture design.}
\label{tab:ablation}
\renewcommand{\arraystretch}{1.2} %
\resizebox{\columnwidth}{!}{%
\begin{tabular}{@{}lcc|cc@{}}
\toprule
Seg.                  & Hair Geom. & Hair-Face Corr. & FID $\downarrow$ & FID-swap $\downarrow$ \\ \midrule
Seg. in $\mathcal{D}$ & \xmark     & cross attn.     & $296.97$         & $296.89$              \\
w/o Seg. loss         & \xmark     & cross attn.     & $10.56$          & $31.97$               \\
w/ Seg. loss          & \xmark     & cross attn.     & $12.15$          & $34.18$               \\
w/ Seg. loss          & \cmark     & --              & $12.11$          & $29.99$               \\
w/ Seg. loss          & \cmark     & concat.         & $10.30$          & $\mathbf{14.95}$      \\
w/ Seg. loss          & \cmark     & cross attn.     & $\mathbf{7.67}$  & $20.56$               \\ \bottomrule
\end{tabular}%
}
\end{table}

\subsection{Latent Space Interpolation}
\begin{figure*}[ht]
    \centering
    
    \begin{tabular}{cc}
      \raisebox{78pt}{\rotatebox[origin=c]{90}{Hair Interpolation}} &
      \includegraphics[width=.94\linewidth]{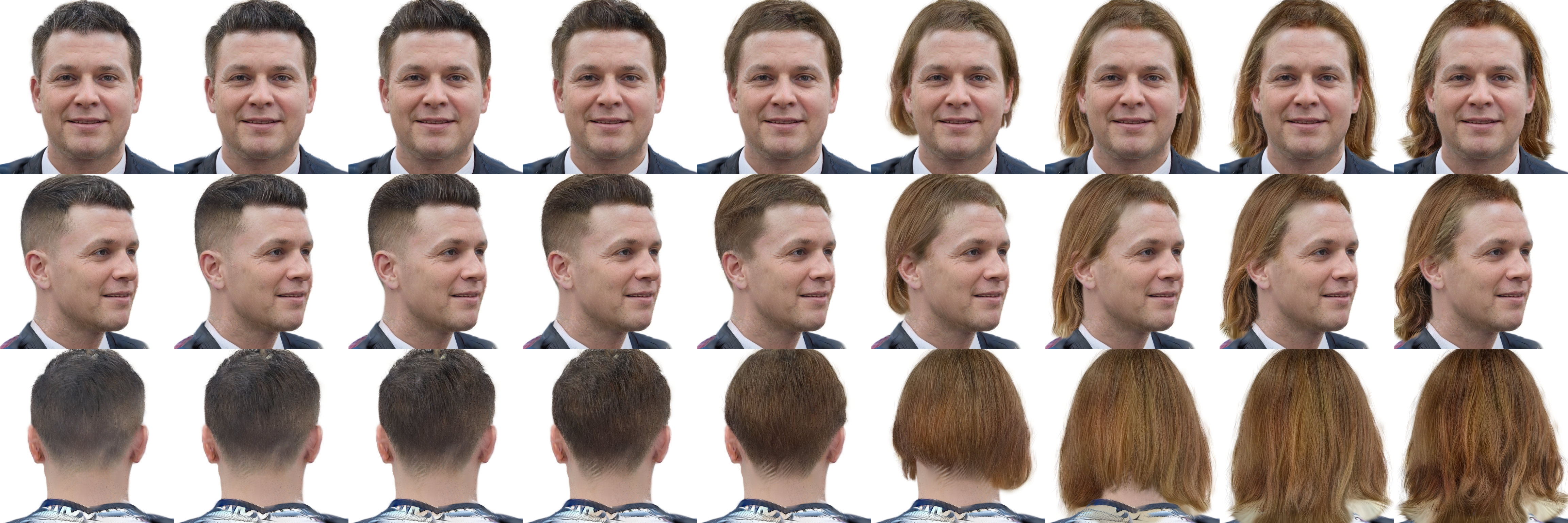} \\
      \raisebox{78pt}{\rotatebox[origin=c]{90}{Face Interpolation}} &
      \includegraphics[width=.94\linewidth]{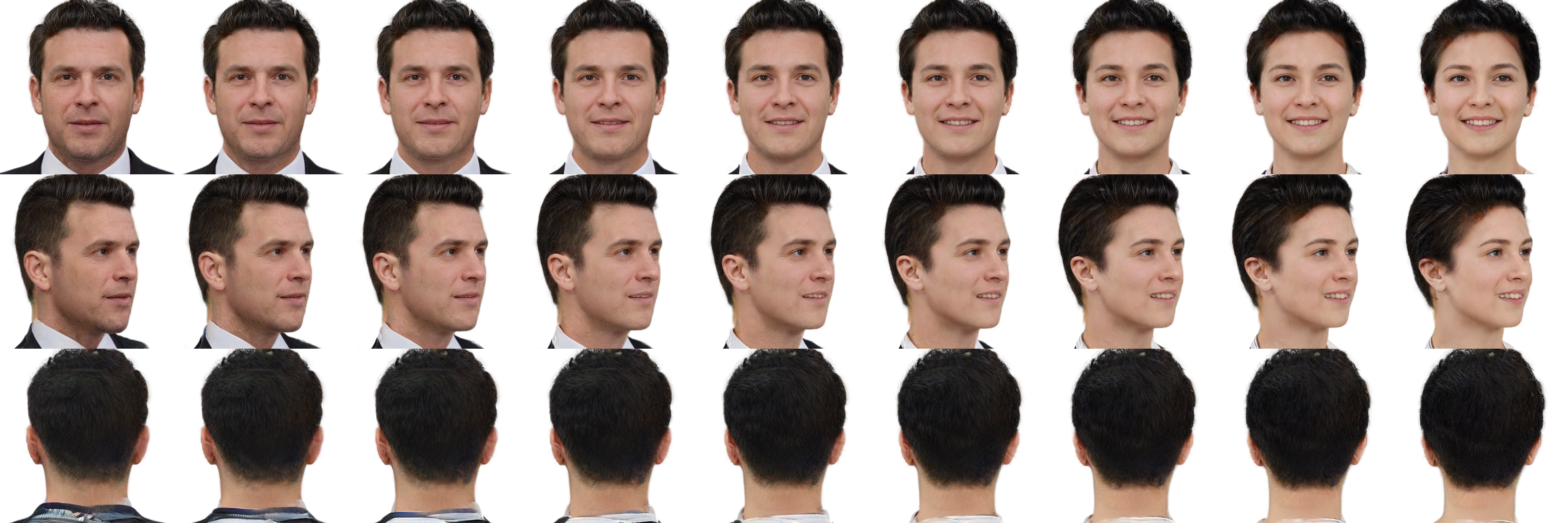} \\
    \end{tabular}
    \caption{\revision{Linear interpolation in the latent space of 3DGH. Top: Interpolation of $\mathbf{w}_{\text{hair}}$ with the facial identity fixed, demonstrating smooth transitions between hairstyles. Bottom: Interpolation of $\mathbf{w}_{\text{face}}$ while keeping the hairstyle constant, illustrating gradual changes in facial features.}}
    \label{fig:lerp}
\end{figure*}

\revision{
\cref{fig:lerp} illustrates how variations in the generated 3D head correspond to interpolations in the latent space of 3DGH. We begin by randomly sampling two pairs of latent codes, which are linearly interpolated to produce intermediate representations. The resulting renderings, arranged from left to right, show a smooth semantic transition between two identities. Leveraging our composable design, we independently interpolate between $\mathbf{w}_{\text{hair}}$ and $\mathbf{w}_{\text{face}}$, enabling disentangled control over hair and facial features.
The seamless transitions and consistent semantic structure across interpolations highlight the continuity and expressiveness of the latent space learned by our model.
}

\section{Discussion}
\label{sec:discussion}

\begin{figure}[t]
  \centering
  \includegraphics[width=\linewidth]{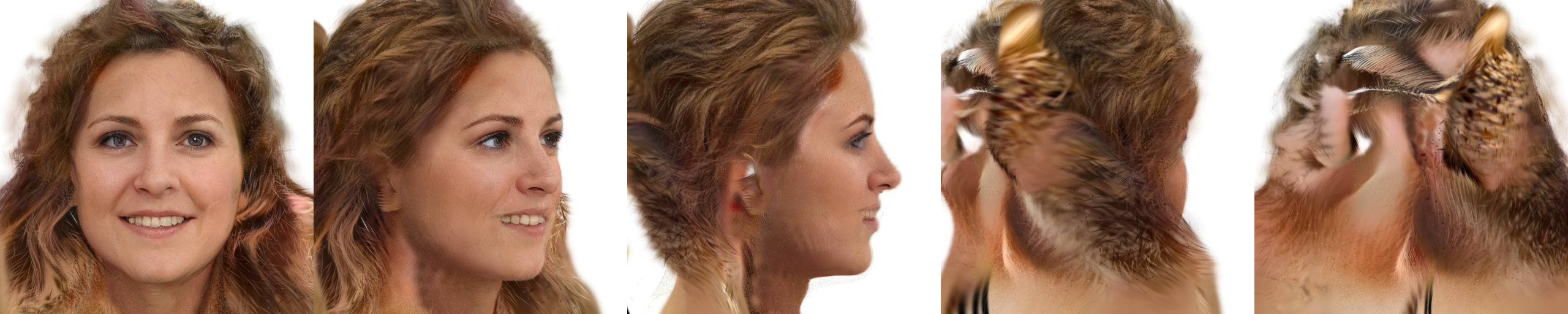}
  \caption{\revision{Failure case illustrating artifacts in the back-view rendering of a long hairstyle.}}
  \label{fig:failure}
\end{figure}

\paragraph{\revision{Limitations and Future Work}}
\revision{
While our method demonstrates strong performance, it still faces several limitations. First, the expressiveness of our model is constrained by the quality and diversity of the training data, which in our case are the images generated by PanoHead~\cite{An_2023_CVPR}. A clear domain gap exists between these synthetic images and in-the-wild images, making certain hairstyles, such as buns and braids, difficult to generate. Addressing this issue necessitates a large-scale dataset of in-the-wild images with comprehensive coverage of frontal and back views, accurate camera calibration, and reliable image alignment. Although works like PanoHead~\cite{An_2023_CVPR} and SphereHead~\cite{li2024spherehead} have made progress in this direction, their in-house training data are not publicly available at this time, and specific processing steps are still needed to calibrate and align back views in the absence of facial landmarks. Therefore, combining real-world multi-view datasets such as RenderMe-360~\cite{pan2023renderme} with synthetic data would be an interesting future work to explore that may alleviate these training data issues.
Second, our model occasionally produces back-view artifacts for long hairstyles that occupy a large portion of the frontal view, as illustrated in~\cref{fig:failure}. We attribute this to our generator’s conditioning on both the latent code and camera pose, following the design choice of EG3D~\cite{chan2022efficient} and PanoHead~\cite{An_2023_CVPR}. Despite using an $80\%$ pose-swapping probability during training, rendering quality degrades when rendering poses differ significantly from the conditioning poses. This limitation is also observed in PanoHead~\cite{An_2023_CVPR} and SphereHead~\cite{li2024spherehead} because full-head image synthesis requires 360-degree rendering. Better conditioning on camera pose may require an advanced architecture as a future work.
Third, our method can produce hollow artifacts, particularly in hair regions, due to the stretching and scaling of Gaussian primitives to represent thin strand structures. While increasing the number of Gaussians may alleviate this problem by providing denser coverage, it would also lead to higher computational cost in terms of model size and training time.
Fourth, while our design choice of the cross-attention layer in~\cref{eq:cross-attn} is motivated by the real-world observation that there exist correlations between ethnic facial features and culturally associated hairstyles, our analysis in~\cref{fig:cfg-scale} does not fully cover various correlations other than gender. This is mainly because gender is the dominant factor in hair-face correlations in the dataset we used. 
Lastly, extending our framework toward animatable 3D avatars is an exciting future direction. By incorporating parametric models such as FLAME~\cite{FLAME:SiggraphAsia2017}, our method could be adapted to support disentangled control over head pose, facial expression, and hairstyle, paving the way for more versatile and customizable 3D avatar generation.
}

\paragraph{\revision{Ethical considerations}}
\revision{
As with other generative models for digital avatars, our method carries potential risks related to misuse (e.g., identity manipulation) and biases. These concerns are partly due to the use of synthetic training data, which may lack sufficient diversity in demographics and hairstyles, limiting the representation of hair-face correlations across different ethnicities (see Fig. 4 in supplemental). To mitigate such issues, we strongly advocate for the responsible use of 3DGH, transparency in its deployment, and the continued development of diverse, representative datasets. We explicitly oppose any use of our work for malicious purposes, including the spread of misinformation or the violation of individual rights.
}

\section{Conclusion}
\label{sec:conclusion}

We introduce 3DGH, a Gaussian-based 3D GAN framework that supports composable hair and face generation. Leveraging multi-view studio capture data, we propose a novel data representation with template-based 3DGS, in which hair Gaussians are rigged to a deformable hair geometry constructed using PCA-based linear blend shapes. This data representation drives the design of our network architecture, which incorporates dual branches to independently generate hair and face Gaussians. A cross-attention mechanism is introduced to model the inherent correlation between hair and face, ensuring coherent and realistic outputs. Our model is then trained using a comprehensive objective that includes adversarial loss, reconstruction terms, and regularization terms designed to stabilize training and facilitate hair-face separation. We evaluate the trained model both qualitatively and quantitatively, demonstrating its superior performance in unconditional full-head image synthesis and composable 3D hairstyle editing.

\bibliographystyle{ACM-Reference-Format}
\bibliography{main}


\end{document}